\documentclass[12pt,preprint]{aastex}

\begin{document}

\title{On the Similarity between Cluster and Galactic Stellar Initial
Mass Functions}

\author{Bruce G. Elmegreen \affil{IBM Research Division, T.J. Watson
Research Center, P.O. Box 218, Yorktown Heights, NY 10598, USA,
bge@watson.ibm.com} }

\begin{abstract}
The stellar initial mass functions (IMFs) for the Galactic bulge,
the Milky Way, other galaxies, clusters of galaxies, and the
integrated stars in the Universe are composites from countless
individual IMFs in star clusters and associations where stars
form.  These galaxy-scale IMFs, reviewed in detail here, are not
steeper than the cluster IMFs except in rare cases. This is true
even though low mass clusters generally outnumber high mass
clusters and the average maximum stellar mass in a cluster scales
with the cluster mass. The implication is that the mass
distribution function for clusters and associations is a power law
with a slope of $-2$ or shallower. Steeper slopes, even by a few
tenths, upset the observed equality between large and small scale
IMFs. Such a cluster function is expected from the hierarchical
nature of star formation, which also provides independent evidence
for the IMF equality when it is applied on sub-cluster scales. We
explain these results with analytical expressions and Monte Carlo
simulations. Star clusters appear to be the relaxed inner parts of
a widespread hierarchy of star formation and cloud structure. They
are defined by their own dynamics rather than pre-existing cloud
boundaries.
\end{abstract}


\section{Introduction}
The initial mass function (IMF) for stars in star clusters is
similar to the integrated IMF in whole galaxies, even though the
integrated IMF is a sum from many clusters which themselves have a
decreasing mass function.  This observed similarity has been used
to suggest that stellar mass is effectively independent of cluster
mass (Elmegreen 2000, 2001), aside from the obvious condition that
a cluster cannot produce a star more massive than itself. Without
such independence, the relatively large number of low mass
clusters would overpopulate the field with purely low mass stars
and tilt the integrated IMF to be steeper than the IMF inside each
cluster.

Kroupa \& Weidner (2003) turned the argument around, suggesting
that the tendency for more massive clusters to contain more
massive stars, which comes from statistical sampling, is enough to
make the summed IMFs from all clusters steeper than the individual
IMFs. Assuming a cluster mass function $dn_{cl}/dM_{cl}\propto
M_{cl}^{-\beta}$ with a slope $\beta=2.2$, they obtained a summed
IMF with a slope of $-2.8$ when the intrinsic cluster IMF has a
slope of about $-2.35$, the Salpeter value (see compilation of
cluster IMFs in Scalo 1998; see review in Chabrier 2003).
Correcting for 100\% binaries steepened the summed IMF in Kroupa
\& Weidner (2003) to a slope of $-3.2$. With the further
assumption that most stars form in clusters, they concluded that
the IMF for whole galaxies should be much steeper than the
Salpeter function. This conclusion has important consequences for
integrated galaxy colors, metallicities, and supernova rates (see
also Goodwin \& Pagel 2005 for a discussion of the supernova
rate).

This steepening of the integrated IMF for galaxies compared to the
IMF for clusters depends sensitively on the cluster mass function
(see Fig. 2 in Kroupa \& Weidner 2003 and Sect. 3 below).  In
particular, the steepening is negligible for $\beta\lesssim2$. To
investigate $\beta$ further, Weidner, Kroupa, \& Larsen (2004)
considered the correlation between maximum cluster luminosity,
$L_{\rm cl,max}$, and galaxy star formation rate, SFR. This
correlation has a lot of dispersion so the slope of the cluster
mass function cannot be well constrained from the data alone.
However, they note that $\beta=2.35$ makes the formation time of a
cluster population independent of the total cluster mass. This
derived power of 2.35 is essentially one larger than the power in
the observed correlation ${\rm SFR}\propto L_{\rm cl,max}^{1.35}$.
To obtain $\beta=2.35$, the maximum cluster mass, $M_{\rm
cl,max}$, was assumed to be proportional to the maximum cluster
luminosity. Such a proportionality requires the maximum-mass
clusters in each galaxy to have the same age, regardless of SFR.
Larsen (2002) showed, however, that a cluster population with a
uniform distribution of ages and a maximum cluster mass (limited
by galaxy size, for example) has a luminosity function that is
steeper than the mass function. In the case he considered, the
luminosity function had a slope of $-2.72$ while the mass function
had a slope of $-2$. The difference arises because the high
luminosity clusters are systematically younger than than the low
luminosity clusters, considering the cluster mass cutoff. Thus the
slope of the cluster mass function derived by Weidner, Kroupa \&
Larsen (2004) should be considered an upper limit. If cluster mass
functions are shallower than $\beta=2.35$, then the difference
between the integrated galaxy IMF and the cluster IMF is smaller
than they suggest.

Weidner \& Kroupa (2005) considered the additional influence of
galaxy size.  Small galaxies, which generally have low total star
formation rates, should have fewer clusters and a smaller maximum
mass for their cluster populations, compared to large galaxies.
Models with this effect and the cluster $\beta=2.35$ suggested
that dwarf galaxies should have steeper integrated IMFs than large
galaxies, with slopes between $-3.1$ and $-3.3$ when individual
cluster IMFs have the Salpeter slope. These authors also returned
to the case where $\beta=2.0$ (their Fig. 8), and again found a
negligible effect on the summed IMF (i.e., a steepening in slope
from the Salpeter value by only 0.1), even in the dwarf galaxy
case.

The cluster mass function has been measured for several galaxies.
Zhang \& Fall (1999) observed $\beta\sim1.95\pm0.03$ and
$\beta=2.00\pm0.08$ for young and old clusters respectively in the
Antennae galaxy, which is the largest sample available for any
galaxy. Another large sample is in recent HST data of M51, where
Gieles et al. (2006) fitted the luminosity function to a mass
function with $\beta=2.0$ and an upper mass cutoff $\sim10^5$
M$_\odot$. For the LMC, the nearest galaxy whose complete cluster
population can be studied, Elmegreen \& Efremov (1997) obtained
$\beta\sim2.0$ using clusters from Bica et al. (1996); Hunter, et
al. (2003) obtained $\beta=2$ to 2.4 in two different ways using
clusters from Bica et al. (1999) with photometry from Massey
(2002); and de Grijs \& Anders (2006) obtained $\beta=1.85\pm0.05$
using the same data as in Hunter et al. but different age
calibrations. Other cluster studies of whole galaxies recently got
similar $\beta$; for example de Grijs et al. (2003) found
$\beta=2.04\pm0.23$ for 147 massive clusters NGC 3310 and
$\beta=1.96\pm0.15$ for 177 clusters in NGC 6745.  Larsen (2002)
considered 6 nearby galaxies and derived luminosity functions with
slopes of 2 to 2.4 but did not derive mass functions; most likely,
the mass functions are shallower (see above).  Evidently,
$\beta\sim2$ is not uncommon for cluster systems. This implies the
IMF on galactic scales should be approximately equal to the
individual cluster IMFs.

These theoretical predictions about the summed IMF should be
compared with observations. Does the integrated IMF in galaxies
have a slope as steep as $-3$? Do dwarf galaxies have steeper IMF
slopes than large galaxies? If not, then how can we understand
intuitively why integrated IMFs are approximately equal to
individual cluster IMFs? In what follows, we begin with a more
detailed discussion of the $\beta=2$ case, and then we discuss the
observations of summed IMFs in Section 3. Section 4 demonstrates
some properties of summed IMFs using Monte Carlo simulations that
include a soft rejection criterion for overly massive stars in a
cluster. This criterion seems appropriate for clouds that form
stars with an efficiency less than one. It also tends to make the
summed IMF more similar to the individual cluster IMFs, even when
$\beta>2$.

\section{Summed IMFs in the $\beta=2$ case: An Intuitive Explanation}

We consider in this section the special case where the cluster
mass function is a power law with decreasing slope $\beta=2$. We
assume that this function applies to both clusters and loose
stellar groupings, and that the stars in each have the same IMF.

Generally, the maximum stellar mass in a cluster, $M_{max}$, found
on average from the equation
\begin{equation}\int_{M_{max}}^\infty n_{s}(M)dM=1\label{basic}\end{equation}
for stellar IMF $n_{s}=n_{s0}M^{-1-x}$, is related to the
normalization factor $n_{s0}$ by the expression
$n_{s0}=xM_{max}^x$ (Sect. 3 considers a more realistic IMF but
gets the same result as in this discussion). Here the notation
corresponds to $x=1.35$ for the Salpeter IMF. The cluster mass is
\begin{equation}M_{cl}=\int_{M_{min}}^\infty
Mn_s(M)dM=n_{s0}M_{min}^{1-x}/\left(x-1\right)=n_{s0}/A\label{mcl},\end{equation}
where $A=(x-1)M_{min}^{x-1}$ is a constant.  The cluster mass
$M_{cl}$ is proportional only to the IMF normalization factor
$n_{s0}$ and the fixed minimum stellar mass, $M_{\min}$. (In this
paper, the notation will denote cluster mass by a subscript
``cl.'') Thus the number of stars in a cluster with a particular
stellar mass range, $M$ to $M+dM$, is directly proportional to the
mass of the cluster:
\begin{equation}
n_s(M|M_{cl})=AM_{cl}M^{-1-x}.
\end{equation}
We use the notation for conditional probability, writing
$n_{s}(M|M_{cl})$ for the IMF inside a cluster of mass $M_{cl}$.

Now consider the IMF for the sum of all clusters, $n_{sum}(M)$.
This is given by the integral over the IMFs for each cluster of
mass $M_{cl}$, weighted by the cluster mass function,
$n_{cl}(M_{cl})=n_{cl,0}M_{cl}^{-\beta}$:
\begin{eqnarray}\nonumber
n_{sum}(M)=\int_{M_{cl}(M_{max}=M)}^{M_{cl,max}}
n_s(M|M_{cl})n_{cl}(M_{cl})dM_{cl}=\\
AM^{-1-x}\int_{M_{cl}(M_{max}=M)}^{M_{cl,max}}
n_{cl,0}M_{cl}^{1-\beta}dM_{cl} \label{nsum}. \end{eqnarray} The
lower limit to the integral is the cluster mass which has a
maximum stellar mass equal to $M$, from equation \ref{mcl},
\begin{equation}
M_{cl}(M_{max}=M)=xM^x/A\label{mmax}.\end{equation} This is a
lower limit to the integral because only this cluster mass and
larger cluster masses are likely to contain a star of mass $M$ (on
average). For $\beta=2$, the integral in equation \ref{nsum} is
\begin{equation}
n_{sum}(M)=An_{cl,0}M^{-1-x}\ln\left(AM_{cl,max}/xM^x\right).\end{equation}
This is essentially the same as the individual cluster IMF because
the logarithmic function varies slowly with stellar mass $M$.

As an illustration of this point, suppose there are 100 clusters
with masses between $10^2$ and $10^3$ M$_\odot$ and 10 clusters
with masses between $10^3$ and $10^4$ $M_\odot$ (which follows
from $\beta=2$). The IMF normalization $n_{s0}$ is proportional to
cluster mass, so we set $n_{s0}=1$ for the first interval and
$n_{s0}=10$ for the second. The summed IMF is the number of
clusters times the IMF of each, which is $n_{s0}M^{-1-x}$, so the
sum is $100M^{-1-x}$ for both intervals. This has the same overall
mass dependence as the IMF in each cluster.

The reason low mass clusters do not tilt the summed IMF to
significantly steeper slopes when $\beta\sim2$ is that, with the
above assumptions (Eq. \ref{basic}), the probability of forming a
star of a particular mass is independent of the cluster mass. The
low $M_{cl}$ bin has a low $M_{max}$ {\it on average} because
$M_{max}$ scales with the cluster mass as $M_{max}\propto
M_{cl}^{1/x}$ from equation \ref{mmax}. But there are 10 times
more clusters with this low cluster mass than in the high mass
bin, and in all of these additional clusters the actual maximum
mass varies a lot around the average.  In fact, there is likely to
be one {\it extreme} example of a low-mass cluster with a maximum
stellar mass that equals the {\it average} maximum mass in the
higher cluster mass bin.  Thus each bin of cluster mass has the
same maximum stellar mass, and the same number of all other
stellar masses.

We demonstrate this point using Monte Carlo simulations in Figure
1, below. A simple argument goes like this. Consider a cluster of
mass $M_{cl}$ which has the nominal or average maximum stellar
mass given by equations \ref{basic} and \ref{mcl},
$M_{max}=\left(AM_{cl}/x\right)^{1/x}$. Now find the maximum
stellar mass that is likely to occur, not in every cluster as
assumed by equation \ref{basic}, but once in some small fraction
$f$ of all the clusters with the same mass $M_{cl}$. This
different maximum, $M_{max,f}$, is given by the equation
\begin{equation}
\int_{M_{max,f}}^\infty n_s(M)dM=f,\end{equation} because the
integral is the number of times the maximum $M_{max,f}$ occurs,
and we are looking for a number that is the fraction of $f$ the
number of total clusters. Solving for $M_{max,f}$ with the same
$M_{cl}$ and therefore the same IMF normalization
$n_{s0}=xM_{max}^x$, we get $M_{max,f}=M_{max}f^{-1/x}$. Now we
ask: what is the cluster mass $M_{cl,f}$ which has $M_{max,f}$ as
an {\it average} maximum mass? This comes from equation \ref{mmax}
and is
\begin{equation}
M_{cl,f}=xM_{max,f}^x/A=xM_{max}^x/\left(Af\right)=M_{cl}/f.
\label{mclf}\end{equation} This result says that the {\it
absolute} maximum stellar mass in $1/f$ clusters of mass $M_{cl}$
is equal to the average maximum stellar mass in one cluster of
mass $M_{cl}/f$. The numerical example above took $f=0.1$. This is
true regardless of the cluster mass function slope $\beta$. It
follows only from the power law nature of the IMF, regardless of
the actual power $x$.  The cluster mass function comes in when we
realize that for $\beta=2$, there are in fact $1/f$ clusters of
mass $M_{cl}$ for every one cluster of mass $M_{cl,f}=M_{cl}/f$.

The summed IMF is equal to the individual IMF also if $\beta<2$,
because then the integral in equation \ref{nsum} depends only on
the upper limit to the cluster mass, which is independent of $M$.

This derivation showing that the slope of the summed IMF is about
equal to the slope of the individual IMF for $\beta\leq2$ is
independent of the IMF itself. Thus it applies to both the IMF
slope for intermediate to high mass stars as well as the shallower
slope for low mass stars.

Observations of galactic-scale IMFs are presented in the next
section. They uniformly support this picture: the summed IMFs from
all the clusters and dissolved clusters that ever lived is about
the same as the average IMF in any one cluster. The implication is
not that star formation samples clouds in a strange way, but
only that the cluster mass function has a slope of $\beta\sim2$,
as observed directly in many cases.

This slope $\beta$ is also suggested by the hierarchical nature of
star formation itself. This structure is commonly observed on
scales ranging from sub-parsec clusters to OB associations to
kpc-size star complexes (see Elmegreen et al. 2006 and references
therein). Very young, embedded clusters show this hierarchy of
stellar positions too, as in Taurus (Gomez et al. 1993), NGC 2264
(Dahm \& Simon 2005), rho Ophiuchus (M. Smith, et al. 2005),
Serpens (Testi et al. 2000), LMC OB association LH 5
(Heydari-Malayeri et al. 2001), and W51 (Nanda Kumar, Kamath, \&
Davis, 2004). Very small stellar systems, consisting of only three
or four stars inside a cluster, can also be hierarchical (e.g.,
Brandeker, Jayawardhana \& Najita 2003). Numerical simulations
showing this effect are in Bonnell, Bate \& Vine (2003): stars
appear to be born with a hierarchical distribution of sub-units. A
stellar grouping presumably loses this structure only when the
stellar ages become comparable to the cluster dynamical time, at
which point the stars move away from their birth positions and mix
together as a result of gravitational forces.  We believe this is
essentially the origin of star clusters: they are the inner mixed
regions of a pervasive hierarchy of stellar birth positions
resulting from gravitational fragmentation and turbulence.

In any hierarchically nested distribution, the number of units
scales with mass as $M^{-2}dM$ because the same total mass is
present at each logarithmic level in the hierarchy (a more general
case is in Elmegreen 2002). Then $\beta=2$ for the sub-units. In
that case, the summed IMFs from all the sub-units in a cluster
equals approximately the IMF of the whole cluster. Otherwise, more
massive clusters would have steeper IMFs, which is contrary to
observations (e.g., Fig. 5 in Scalo 1998). Thus the equality
between summed IMFs and individual IMFs is a common occurrence
{\it inside} the cluster environment as it is outside the cluster
environment, in whole galaxies. All of these regions are part of
the hierarchy of star formation.

Numerical simulations of cluster formation suggest a far more
complicated picture of how the IMF arises (see review in Bonnell,
Larson, \& Zinnecker 2006). Subclumps have their own mass
segregation and their own enhanced accretion to the most central
stars (Bonnell, Vine \& Bate 2004). Then they blend together over
time, with continued growth of the most massive stars. However,
these processes appear to be a microcosm of what happens on larger
scales: star clusters form clustered together, each with their own
cloud core, mass segregation, and IMF, each accretes from the
surrounding parts of the ISM, and they sometimes blend together
over time into bigger clusters. Other times they blend into OB
associations after locally dispersing. In either case, they add up
to give about the same IMF in the composite region as they had
individually. In terms of the physical processes at work, there is
nothing random about this except for the smallest details in the
initial conditions that get amplified by turbulence and
self-gravity. Still, the basic equations for the distribution
functions should be the same as in our simple models, as should
the requirement that $\beta\sim2$ to make the sum of all the IMF
subunits equal to the observed IMF whole.

\section{Observations of Galaxy-Integrated IMFs}

The IMF for a galaxy is the sum of the IMFs of the individual
star-forming regions and the IMFs of all the captured dwarfs and
other extragalactic systems. This sum includes both field stars
and the stars remaining in clusters. It cannot be observed
directly using star counts unless the star formation history is
known.  This differs from the situation in globular clusters and
other clusters where the stars presumably formed in a single
burst. For galaxies, the integrated IMF may be determined, subject
to certain assumptions, from broad-band colors, H$\alpha$
equivalent widths, star counts of low-mass (unevolved) stars, and
the relative abundances of various elements. Essentially all of
the observations suggest the integrated IMF may be approximated by
a power law at intermediate to high mass with a slope comparable
to that of the Salpeter function ($dN/dM\propto M^{-\alpha}$ for
$\alpha\sim2.35$) or slightly steeper (e.g., $\alpha\sim-2.7$),
and a flattening below a mass of $\sim1$ M$_\odot$ or slightly
less. Elliptical galaxies and clusters of galaxies may have formed
with a slightly flatter IMF, which also appears to be
characteristic of the Universe as a whole. Little is known about
the integrated IMF below $0.1$M$_\odot$. Here we review the recent
observations of this summed IMF because there is no comparable
review in the literature.

For the local part of the Milky Way, star counts and metallicities
suggest that the slope of the intermediate to high mass part of
the IMF is comparable to or slightly steeper than the Salpeter
slope (Scalo 1986; Rana 1987; Tsujimoto et al. 1997; Thomas,
Greggio \& Bender 1998; Boissier \& Prantzos 1999; Gratton et al.
2000; de Donder \& Vanbeveren 2003; Portinari et al. 2004).
Carigi, et al. (2005) require an IMF slope of $-2.7$ to explain
the C/H, N/H, and O/H ratios.  This slope is somewhat difficult to
constrain because it depends on the upper stellar mass limit, the
star formation history, gas accretion rate, stellar migration,
stellar evolution models, and other things.

In the Milky Way bulge, several of these uncertainties are not so
important (e.g., age variations, migration). There, the integrated
IMF for $0.15<M/{\rm M}_\odot<1$ is similar to that of globular
clusters (Holtzman et al. 1998; Zoccali et al. 2000). The
galaxy-wide IMF in the local dwarf Spheroidal Ursa Minor is also
indistinguishable from the IMFs of Milky Way globular clusters
(Feltzing, Gilmore \& Wyse 1999). Similarly, the IMF in the Draco
Dwarf Spheroidal is like that in the old cluster M68 (Grillmair et
al. 1998). If stars typically form in clusters that subsequently
disperse to make a bulge or a dwarf galaxy, then these three
examples suggest directly that the summed IMF from a population of
clusters approximately equals the IMF of each cluster.

Other nearby dwarf and Irregular galaxies have IMFs with slopes
similar to the Salpeter value. Crone et al. (2002) find from
models of color-magnitude diagrams that the IMF in the BCD galaxy
UGCA 290 is approximately Salpeter for intermediate mass stars; a
slope as steep as $-3$ overproduces the faint blue main sequence
relative to the brightest supergiants.  Annibali et al. (2003)
determined the range of acceptable IMFs from color-magnitude
models for NGC 1705. In Region 7 of that galaxy, the Salpeter
slope or slightly flatter ($\alpha=-2.35$ to $-2$) for $M>6$
M$_\odot$ gave an acceptable fit -- one steeper produced a
luminosity function that was too steep for the most recent burst.
In Region 6 of NGC 1705, the preferred IMF slope was Salpeter or
possibly steeper, with a limit at about $-2.6$. The dwarf
irregulars DDO 210 and DDO 3109 have approximately Salpeter IMFs
at intermediate mass as well (Greggio et al. 1993). In NGC 1309, a
slightly shallower IMF was preferred ($-2.2$) to match the observed
proportion of main sequence stars to evolved stars; a steeper IMF,
including even the Salpeter IMF, overproduced evolved stars with
moderate mass. NGC 1569, a starbursting dwarf, the IMF was Salpeter
or slightly flatter (Angeretti et al. 2005), as it was in NGC 6822
(Marconi et al. 1995).  NGC 6822 was also studied more recently by
Carigi, Col\'in \& Peimbert (2005) using models of chemical
evolution. Reasonable results were obtained with the Kroupa, Tout
\& Gilmore (1993) IMF, which has a slope at intermediate to high
mass of $-2.7$. They suggest, however, that the upper limit to the
stellar mass should be lower than in Kroupa, Tout \& Gilmore,
namely $\sim60$ M$_\odot$ rather than 80 M$_\odot$.

The IMFs in Wolf-Rayet galaxies were studied by Fernandes et al.
(2004) using the ratio of Wolf-Rayet stars to O-type stars
obtained from spectra.  The low metallicity galaxies of this type
were fitted to an IMF with a slope between $-2$ and $-2.35$ in a
short burst of star formation. The high metallicity galaxies were
found to have either a steeper slope in the bursting models, or
the Salpeter slope ($-2.35$) in more extended bursts. However, the
ratio of Carbon to Nitrogen Wolf Rayet stars supported the
extended burst models. Thus the preferred IMFs were never steeper
than the Salpeter slope in that study.

These observations of dwarf galaxies support the proposal outlined
in the Introduction that the IMF from each cluster equals
approximately the IMF of the whole stellar population. This
contradicts the basic proposal about summed IMFs in Kroupa \&
Weidner (2003), and it also contradicts the models in Weidner \&
Kroupa (2005) which suggest that small galaxies should have
steeper IMFs than large galaxies. The basis for Weidner \&
Kroupa's 2005 model was that smaller galaxies, with very low star
formation rates, should have a small number clusters and therefore
a small mass for the most massive cluster, from statistical
sampling.  Then, if the maximum stellar mass depends on the
cluster mass, the formation of high mass stars should be
relatively rare.  However, even dwarf galaxies can contain super
star cluster with very massive stars in them (Billett, Hunter, \&
Elmegreen 2002). Massey, Johnson, \& Degioia-Eastwood (1995) also
noted that the maximum masses for stars are about the same in the
Milky Way, the LMC and the SMC, despite the size differences among
these three galaxies.

One exception to the near uniformity of IMF slopes for dwarf
galaxies is in the main body of the BCD galaxy I Zw 18, which
apparently has a flat IMF with a slope of $\sim-1.5$ (Aloisi,
Tosi, \& Greggio 1999). This conclusion was based on
color-magnitude diagrams obtained with the Hubble Space Telescope
and is clearly different from that in all of the other dwarfs
studied by this group using the same methods. The origin of this
discrepancy has not been explained, but in any case, the observed
slope is not steeper than the nominal cluster IMF.

The IMF in the Large Magellanic Cloud has been determined for many
regions. Holtzman et al. (1997) observed a field region for stars
between $\sim0.6$ M$_\odot$ and 1.1 M$_\odot$ and derived an IMF
slope between $-2.0$ and $-3.1$. Parker et al. (1998) determined
IMFs slopes for massive stars in field regions and got
$-2.80\pm0.09$. Selman \& Melnick (2005) observed the LMC field
outside the star formation regions near 30 Dor and found an IMF
slope of $-2.38\pm0.04$ for $7<M/{\rm M}_\odot<40$. This is
essentially the same as the Salpeter IMF.  Selman \& Melnick
pointed out that this field slope is slightly steeper (by 0.12)
than the IMF in the nearby super star cluster, NGC 2070, in
agreement with predictions about field-star IMF steepening by
Kroupa \& Weidner (2003) for $\beta=2$.  Higher $\beta$ would not
give this result. Steeper field IMFs for the LMC will be discussed
below.

Large galaxies have integrated IMF slopes comparable to the
Salpeter value. Fifty OB associations summed together in M31 were
found to have an IMF slope of $-2.59\pm0.09$ using a combination
of space-based uv and HST photometry, 2MASS photometry at JHK
bands, and I band photometry (Veltchev, Nedialkov, \& Borisov
2004). The stellar masses ranged from 8 M$_\odot$ to 100
M$_\odot$. Kennicutt and collaborators measured the integrated
IMFs in several hundred nearby galaxies of various Hubble types
using optical and H$\alpha$ surface brightnesses (Kennicutt 1983;
Kennicutt, Tamblyn \& Congdon 1994; Bresolin \& Kennicutt 1997;
Bresolin et al. 1998).  The IMF did not vary with Hubble type or
galaxy luminosity, and was usually fit with a power law at
intermediate to high mass with a slope of $\sim-2.5$.

The summed IMF for all of the stars in clusters of galaxies has
been studied in detail using constraints from broad-band colors
and the metallicities of the galaxies and the intergalactic
medium. Metallicity is sensitive to the mass fraction of high mass
stars and $\alpha$ abundance ratios are sensitive to the ratio of
intermediate to high mass stars. The general consensus seem to be
that the IMFs in galaxy clusters had to be flatter than Salpeter
at some time in the past, when the metallicity of the
intergalactic medium was established. Then it probably steepened
to the Salpeter slope in recent times. Renzini et al. (1993) fit
the Fe abundance in clusters with a Salpeter IMF in the quiescent
star forming phases and a flatter IMF in the bursts. Loewenstein
\& Mushotsky (1996) fitted elemental abundances in four rich
clusters of galaxies with an IMF flatter than Salpeter (the
suggested slope was in the range $-1.7$ to $-2$ to get enough Fe).
Chiosi (2000) suggested the characteristic (turnover) mass in the
IMF got larger with redshift, as did Moretti, Portinari, \& Chiosi
(2003), who fitted the intra-cluster metallicities with
proportionally more massive stars at high redshift and in more
massive galaxies. Tornatore et al (2004), Romeo et al. (2004),
Portinari et al. (2004), and Nagashima et al. (2005a) also
suggested the intra-cluster abundances came from a top-heavy IMF
in the past, particularly in the most massive and most bursting
galaxies. Pipino \& Matteucci (2004) and Nagashima et al. (2005b)
suggested a slightly top-heavy IMF for massive elliptical
galaxies. All of these studies concluded that the composite IMF
was Salpeter or slightly flatter than Salpeter when galaxy
clusters formed. Still, a Salpeter IMF may be possible if a
significant fraction of intra-cluster Fe comes from Type II
supernovae (Wyse 1997). Simulated color-magnitude diagrams of
ground-based U,V observations of the nearby elliptical galaxy NGC
5128 suggest a Salpeter IMF, with a limit no steeper than $-2.6$
(Rejkuba, Greggio \& Zoccali 2004). There is no evidence that the
integrated IMFs in galaxy clusters are significantly steeper than
Salpeter, even though galaxies and clusters of galaxies formed
from the sum of countless dispersed star clusters.

On a larger scale, Baldry \& Glazebrook (2003) fitted the local
luminosity density and star formation history of the Universe with
population synthesis models and obtained an average IMF slope of
$-2.15\pm0.2$ for all galaxies combined. The upper limit to the
slope was $-2.7$. Lacey et al. (2005) modelled galaxy formation
and matched the observed UV galaxy luminosity function over a wide
range of redshifts with a solar neighborhood IMF that becomes
top-heavy in starbursts. Calura \& Matteucci (2004) fitted the
relative abundance of elements in galaxies of various types and
found the Salpeter IMF was the best choice for a universal IMF.

Evidently, most observations of galaxies suggest the composite IMF
is not significantly steeper than the Salpeter IMF, even though
these systems are the combinations of many small clusters and
associations, each of which probably has the Salpeter IMF on
average, like local clusters and associations. There is no
systematic steepening of the IMF in composite fields.

Nevertheless, several observations suggest the IMF in some regions
is significantly steeper than the Salpeter slope. Gouliermis,
Brandner \& Henning (2005) measured the IMF in a field region near
the giant shell LMC4. They obtained a slope of $\sim-6$ for
$0.9<M/{\rm M}_\odot<2$ and $\sim-3.6$ for $0.9<M/{\rm
M}_\odot<6$.  These slopes at low mass agree with those found by
Massey (2002) for high mass stars. There are other galaxies where
steep IMFs have been reported. The high mass-to-light ratio in the
disks of low surface brightness galaxies was taken as evidence for
an IMF slope of $-3.85$ by Lee et al. (2004). Also, Zackrisson, et
al. (2004) suggested that the IMF slope is $-4.5$ in the red
stellar halos around BCD galaxies (Bergvall \& \"Ostlin 2002) and
in the red halos of stacked images of edge-on disks (Zibetti,
White \& Brinkmann 2004). These steep slopes can be explained by
the Kroupa \& Weidner (2003) model with $\beta>2.3$, but there are
other explanations for steep slopes in these particular regions
also (Elmegreen 2004; Elmegreen \& Scalo 2006).

\section{Monte Carlo Models}

The difference between the IMF of a single cluster and the summed
IMFs of many clusters can be seen from Monte Carlo simulations
where star and cluster masses are randomly chosen from
distribution functions (e.g., Weidner \& Kroupa 2003, 2005, 2006).
This is only an approximation meant to simulate the diverse
processes happening in the cluster environment. Cluster
simulations suggest a correlation between cluster mass and star
mass, possibly because of definite physical processes (e.g.,
Bonnell, Vine \& Bate 2004). These are likely to produce
interesting variations around the average condition, some of which
may correlate with environment in ways that are not considered by
these simple models. Still, after real clusters disperse and the
stars mix with other dispersed stars, the summed IMF should depend
on the cluster mass function in the manner suggested by these
simple models, and it should depend on the range of cluster mass
where each star is likely to form.

Here we discuss Monte Carlo calculations like those in Weidner \&
Kroupa, but with an important difference. We select a cloud mass
from a mass distribution function and then select stars until the
total stellar mass exceeds an efficiency factor $\epsilon$ times
the cloud mass. This has the same effect of building up clusters
with a pre-set cluster mass function (equal to the cloud mass
function), but the rejection of stars is soft.  This difference is
important because when stars build up in a cloud core, their
feedback on the gas gets stronger until the gas is expelled and
star formation stops. There should not be a sudden point at which
all star formation stops, but a gradual transition. During this
transition, it is likely that low mass stars continue to form, as
there is generally a lot of gas left over in the cloudy debris.
There may not be enough gas to form high mass stars though. In the
simulation, we include in the cluster any selected star that fits
in the remaining gas reservoir, and then stop the selection
process after the efficiency exceeds the assumed value. This
procedure differs significantly from that in Weidner \& Kroupa
(2003, 2005) for low mass stars. Weidner \& Kroupa stopped the
selection process when the total stellar mass exceeded the pre-set
cluster mass, without regard to the mass of the final excluded
star. Thus their exclusion process operated on stars of all
masses, and their summed IMFs in the $\beta>2$ cases were steeper
for all masses. Our exclusion process operates only on stars more
massive than $1-\epsilon$ times the cloud mass, so our summed IMF
is steepened only for stars more massive than
$\left(1-\epsilon\right)$ times the minimum cloud mass. If stellar
clustering continues all the way down to single stars (Kiss et al.
2006), then this difference is not important. But if star clusters
have a minimum mass of $\sim30$ M$_\odot$ (Lada \& Lada 2003),
then the difference in these two methods is important regardless
of $\beta$. We use this procedure to illustrate a second way in
which the summed IMFs in galaxies can be comparable to the
individual cluster IMFs, distinct from the condition mentioned
previously that $\beta\sim2$. However, even without this
assumption it will be clear from the high mass part of the Monte
Carlo IMF that the $\beta\leq2$ case gives a summed IMF
indistinguishable from the individual IMF.

Our model stellar IMF is a power law at intermediate to high mass
with a generic slope of $-2.5$, i.e., close to the Salpeter value
of $-2.35$ but not intended to match any particular observation.
The IMF turns over at the mass $M_t=0.5$ M$_\odot$ and rises as
$M^{-1}$ below that (i.e., it becomes flat on a log-log plot):
\begin{equation}
n(M_s)dM_s=n_{0s}M^{-2.5}\left(1-e^{-\left[M/M_t\right]^{1.5}}\right)
dM_s.\label{imf}
\end{equation}

Clusters are considered to form inside gas clouds with an
efficiency $\epsilon$. The cloud mass function is a power law
proportional to the final cluster mass function:
\begin{equation}
n_{cld}\left(M_{cld}\right)dM_{cld}=n_{0,cld}M_{cld}^{-\beta}dM_{cld}.
\label{ncl}
\end{equation}
Here $M_{cld}$ represents a cloud mass.  The average stellar mass
inside each cloud is $M_{cl}=\epsilon M_{cld}$. The efficiency is
constant for each cluster in the summed IMF, so the final cluster
mass distribution is the same power law as the cloud distribution.

The lower and upper mass limits to the IMF are taken to be
$M_{min}=0.01$ M$_\odot$ and $M_{max}=150$ $M_\odot$ (following
Weidner \& Kroupa 2004; Oey \& Clarke 2005; Koen 2006). For
clouds, $M_{cld,min}=10$ M$_\odot$ or 1 $M_\odot$ in two cases,
and $M_{cld,max}=10^6$ M$_\odot$. The lower limit to the cloud
mass will have an effect on the summed IMF, but the upper limit is
sufficiently high that sampling rarely selects such a massive
cloud, and its effect is small.

The selection process works as follows. First a cloud mass
$M_{cld}$ is randomly chosen from the distribution function
$n_{cld}$ and then stellar masses $M$ are randomly chosen for this
cloud from $n_s(M)$. The running sum of stellar masses in this
cloud is $\Sigma_M$. The remaining mass available for more stars
is $M_{cld}-\Sigma_M$.  If the most recent star chosen has a mass
less than $M_{cld}-\Sigma_M$, then there is room for it in the
cloud and it is added to the list of cluster stars. If the most
recent star has a mass exceeding this value, then the star is not
kept and another star is chosen. This is how stars are rejected.
When the running sum exceeds $\epsilon M_{cld}$, the selection of
stars for that cloud ends and another model cloud is chosen. We
continue in this way until the total number of stars in all
clusters is $10^8$.

We are interested in the summed stellar mass distribution from all
of the clusters, the summed IMF of the rejected stars, and the
dependence of maximum stellar mass on cloud mass.  For this latter
quantity, we evaluate both the average maximum stellar mass for
clouds within a range of values, and the absolute maximum stellar
mass ever chosen for a particular cloud mass range. For the first
quantity we find the maximum stellar mass for all clouds in a
range of cloud masses and average these maxima together. This
average is always less than the second quantity, which is the
absolute maximum stellar mass found in the same clusters.  We
showed in Section 2 that the average maximum stellar mass per
cluster increases in direct proportion to the cluster mass,
independent of $\beta$, whereas the absolute maximum stellar mass
for many clusters within a range of cluster masses is independent
of cluster mass when $\beta=2$.

Figure 1 shows these quantities for $\beta=2$ and $\epsilon=0.1$,
0.3, and 0.9.  In the bottom panel, the lower of the two parallel
curves is the individual cluster IMF from equation \ref{imf} (with
arbitrary vertical shift). The upper curve is actually three
curves superposed that show the summed IMFs for the three values
of $\epsilon$; they overlap nearly exactly. The bottom three
curves are the mass functions for the rejected stars (which are
not necessarily the last stars chosen in a cluster). The top panel
shows the theoretical prediction for the average maximum stellar
mass versus cluster mass, obtained by numerical integration over
equation \ref{imf}. Other curves are the Monte Carlo results for
maximum star mass labelled by $\epsilon$. The smallest cloud mass
is 10 M$_\odot$ in these models, so the smallest cluster masses
are 1, 3, and 9 M$_\odot$ after multiplying by $\epsilon$. Also
shown in the top panel is the absolute maximum stellar mass for
factor-of-ten intervals in cluster mass (the plus, cross, and
circle marks). The line marked ``$\epsilon\times$Cloud Mass'' is
the nominal cluster mass. The absolute maximum stellar mass
exceeds the nominal cluster mass in low-mass clouds when a massive
star is chosen early, before numerous low mass stars deplete the
cloud mass.  The absolute maximum stellar mass cannot exceed the
cloud mass.

Figure 1 shows that the summed IMF is barely steeper than the
cluster IMF for $\beta=2$. The difference in slope is $\sim0.1$
for $M>10$ M$_\odot$. The efficiency $\epsilon$ does not matter
much for this result. Only high mass stars are rejected,
$M>\left(1-\epsilon\right)M_{cld}$ for $M_{cld}=10$ M$_\odot$. In
the top panel, the average maximum stellar mass follows the
prediction, as does the absolute maximum, which is about constant
except for the lowest-mass clusters.  This constancy is the
primary reason why the summed IMF is so similar to the individual
IMF even when there are far more low mass clusters than high mass
clusters. The {\it average} maximum stellar mass per cluster
increases with cluster mass, but the {\it absolute} maximum
stellar mass does not except at very low $M_{cld}$.

Figure 2 shows the summed IMFs in the middle panel, the rejected
IMFs in the bottom panel, and the cluster mass functions in the
top panel for four values of $\beta$, a minimum cloud mass of 10
M$_\odot$, and $\epsilon=0.3$. Also shown in the bottom two panels
is a case with $M_{cld,min}=1$ M$_\odot$, $\beta=3$, and
$\epsilon=0.3$ (dashed lines), and another case with
$M_{cld,min}=10$ M$_\odot$, $\beta=3$, and $\epsilon=0.99$ (dotted
lines). The first shows the effect of minimum cloud mass, which
determines the point in the summed IMF where the steep slope
begins. The second shows the effect of efficiency $\epsilon$. When
the slope of the summed IMF is steep, the number of rejected stars
is large, even comparable to the total number of stars in the
$\beta=2$ case.  Note, however, that when $\beta\leq2$, the
fraction of stars that are rejected is small for all stellar
masses, $<20$\%, as shown also in the bottom of Figure 1. This is
why the summed IMF is about the same as the individual cluster IMF
for small $\beta$. That is, stars with masses comparable to or
larger than the cloud mass are rejected, but their number is
negligible compared to the number of stars that are kept.

The bottom two panels of Figure 2 show how the slope of the summed
IMF follows the slope of the cluster mass function. This
dependence may be seen from solutions to equation \ref{nsum}.
Least squares fits to the summed IMF slopes above 10 $M_\odot$ in
Figure 2 give values of $-1.5$, $-1.6$, $-2.1$, and $-2.6$ for
$\beta=1.5$, 2, 2.5, and 3 when $M_{cld,min}=10$ M$_\odot$ and
$\epsilon=0.3$. When $M_{cld,min}=1$ M$_\odot$, $\beta=3$, and
$\epsilon=0.3$ (dashed curve), the slope is -2.8 above 1
M$_\odot$. When $M_{cld,min}=10$ M$_\odot$, $\beta=3$, and
$\epsilon=0.99$ (dotted curve), the slope is -2.5 above 10
M$_\odot$. The scatter in these slopes is on the order of
$\pm0.05$.

The figures confirm the analytical results of Section 2 that the
summed IMF from a population of clusters is indistinguishable from
the individual cluster IMF when $\beta\leq2$.  The absolute
maximum stellar mass is independent of the cluster mass even
though the average maximum stellar mass increases with cluster
mass.  These results follow from the model because the fraction of
stars that are rejected is negligibly small in the $\beta\leq2$
cases.

\section{Conclusions}

Observations indicate that galactic bulges, whole galaxies of
various types, clusters of galaxies, and the combined populations
of stars in the Universe, all have composite IMFs with a slope
close to the Salpeter value. The IMF appears to be slightly
steeper than this in the Solar neighborhood and in a few other
regions, and to have been slightly flatter in the past in massive
or star-bursting galaxies that populate galactic clusters.
Significantly steeper IMFs have been found in the field regions of
the LMC and inferred from the colors and luminous densities of Low
Surface Brightness galaxies and edge-on galaxy halos. For most
galaxies, however, there is no evidence that composite IMFs are
significantly steeper than individual IMFs observed in clusters
and OB associations. If most stars form in such clusters and
associations, then this near-equality of IMFs implies that the
cluster and association mass distribution function is close to a
power law with a negative slope of $\beta=2$ for linear intervals
of mass.

Stars form with no apparent physical connection between their mass
and the resulting cluster mass. There is a statistical connection
from the size of sample effect, with more massive clusters forming
more massive stars on average, but this statistical connection has
no obvious influence on the summed IMF from many clusters when
$\beta=2$.  The probability that random sampling attempts to
produce a star more massive than a cloud is negligible for
$\beta=2$. There is no correlation between the maximum stellar
mass in a collection of similar clusters and either the cluster
mass or the maximum average mass per cluster. Thus 100 Taurus-size
clouds should be able to produce the same maximum stellar mass and
the same IMF overall as a single cloud with 100 times the mass
(regardless of $\beta$). This was shown analytically in Section 2
and with Monte Carlo simulations in Figure 1. The prediction has
not been confirmed observationally, but such a confirmation could
be subtle. The Orion cloud, for example, should appear to be
composed of 100 Taurus-like smaller clouds when the density
distribution is viewed with sufficiently high resolution and
sensitivity.

These results fit in well with a model in which all clouds and
young stars form in hierarchical patterns. Observations of many
types suggest these patterns extend from kpc scales down to the
scale of individual clusters. There is also tentative evidence for
a continuation down to individual stars when the cluster
environment has not mixed the stars together. In that case,
neither clusters nor clouds are well-defined entities. They are
defined more by observational selection and the dynamical
processes following star formation than by the physics of
individual star formation, which operates even in the pre-mixed
cluster sub-units. What we perceive to be a cluster is most likely
the mixed inner region of the hierarchy of young stars, with a
size given by the approximate equality between dynamical time and
age. In such a model, $\beta=2$ automatically and stars are not
limited by pre-defined cluster masses because there is no such
thing as a pre-defined cluster.

{}

\clearpage
\begin{figure}
\epsscale{0.7} \plotone{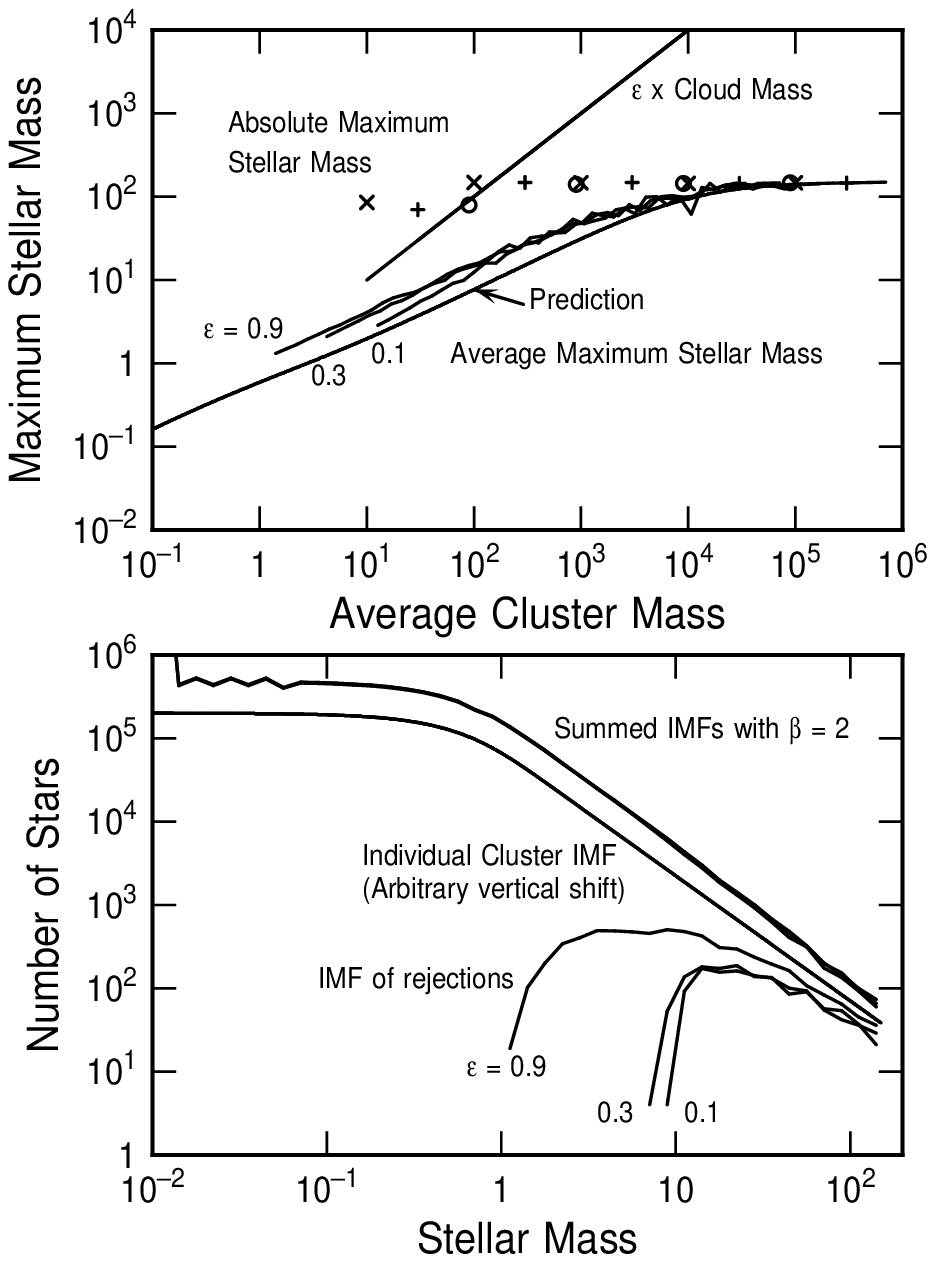} \caption{Monte Carlo IMF models
with a cluster mass function slope $\beta=2$. The bottom panel
shows the individual cluster IMF that is input to the model, and
it shows three summed IMFs for different star-formation
efficiencies $\epsilon$; these three summed IMFs are nearly
identical to each other and their curves overlap. The IMFs of the
rejected stars, which are those too massive to fit in the
remaining gas of the cloud, are at the bottom. The top panel shows
the correlation between the average maximum stellar mass in a
cluster and the cluster mass (increasing curves). The absolute
maximum stellar masses for all clusters, in bins spaced by a
factor of 10 in cluster mass, are shown by symbols centered in
their mass bins. The cases $\epsilon=0.1$, 0.3, and 0.9 are
represented by ``x,'' plus, and circle. The summed IMF for this
$\beta=2$ case is nearly identical to the individual cluster IMF
even though the average maximum stellar mass increases with
cluster mass. This IMF similarity is the result of the
near-constant absolute maximum stellar mass, which is a general
property of power law cluster IMFs, and the $\beta=2$ distribution
for clusters.}\label{fig:imf3}\end{figure}
\newpage

\begin{figure}
\epsscale{0.5} \plotone{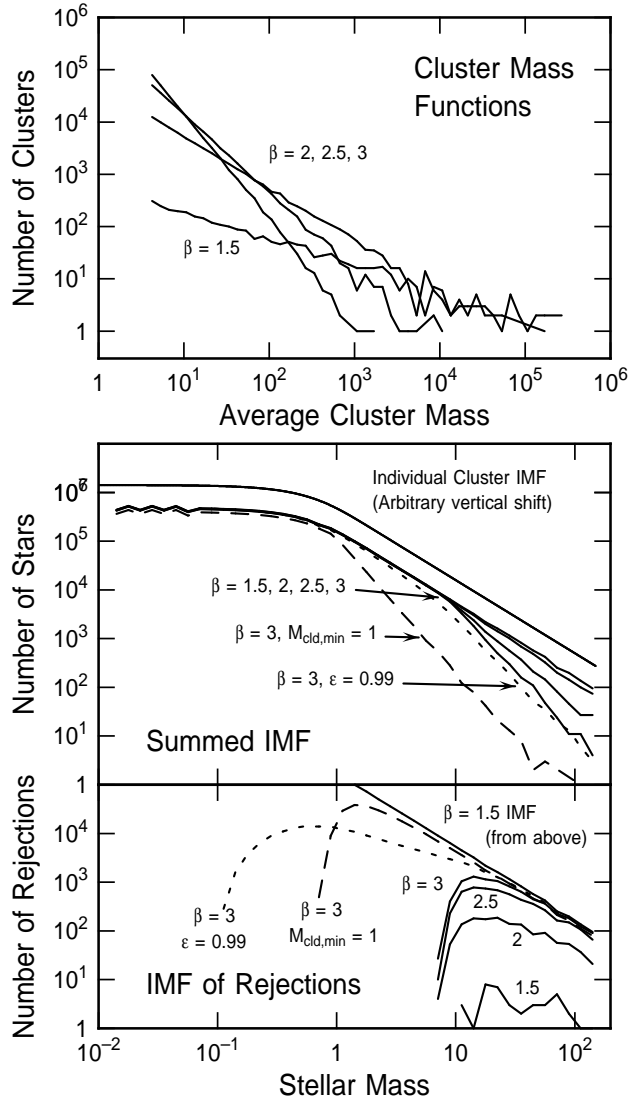} \caption{Summed IMFs for cluster
populations with different cluster mass function slopes $\beta$.
The minimum cloud mass is 10 M$_\odot$ for all curves except the
one indicated, where it is 1 M$_\odot$.  The star formation
efficiency $\epsilon$ is 0.3 for all curves except the comparison
case with $\epsilon=0.99$. The summed IMF in the middle panel gets
steeper with steeper cluster mass function, but agrees with the
individual cluster IMF for $\beta\leq2$. The summed IMF steepening
begins at the minimum cloud mass for producing a cluster,
depending slightly on $\epsilon$. The bottom panel confirms that
the rejection IMF becomes large at the stellar mass where the
summed IMF begins to get steep. The randomly selected cluster mass
functions used for these summed IMFs are shown in the top panel
for the four cases with $\epsilon=0.3$ and $M_{cld,min}=10$
M$_\odot$.  These cluster masses are on average $\epsilon$ times
the cloud masses.}\label{fig:imf3f2}\end{figure}

\end{document}